\documentclass[aps,twocolumn]{revtex4-1}
\usepackage{latexsym}
\usepackage{amsmath}
\usepackage{amssymb}
\usepackage{amsfonts,dsfont}
\usepackage{subfigure}
\usepackage[english]{babel}
\usepackage[cp1250]{inputenc}
\usepackage[OT4]{fontenc}
\usepackage[pdftex]{graphicx}

\begin{document}

\title{Stabilizing effect of tip splitting on the interface motion}
\author{Michal Pecelerowicz and Piotr Szymczak }\email{Piotr.Szymczak@fuw.edu.pl}
\affiliation{Institute of Theoretical Physics, Faculty of Physics, University of Warsaw,  
Pasteura 5, 02-093 Warsaw, Poland}

\begin{abstract}

Pattern-forming processes, such as electrodeposition, dielectric breakdown or viscous fingering are often driven by instabilities. Accordingly, the resulting growth patterns are usually highly branched, fractal structures. However, in some of the unstable growth processes the envelope of the structure grows in a highly regular manner, with the perturbations smoothed out over the course of time. In this paper, we show that the regularity of the envelope growth can be connected to small-scale instabilities leading to the tip splitting of the fingers at the advancing front of the structure. Whenever the growth velocity becomes too large, the finger splits into two branches. In this way it can absorb an increased flux and thus damp the instability. Hence, somewhat counterintuitively, the instability at a small scale results in a stability at a larger scale. The quantitative analysis of these effects is provided by means of the Loewner equation, which one can use to reduce the problem of the interface motion to that of the evolution of the conformal mapping onto the complex plane. This allows an effective analysis of the multifingered growth in a variety of different geometries. We show how the geometry impacts the shape of the envelope of the growing pattern and compare the results with those observed in natural systems.
\end{abstract}

\maketitle 


\section{Introduction}

A variety of natural growth processes, including viscous fingering, solidification, and electrodeposition, can be modeled in terms of Laplacian growth. Laplacian growth patterns are formed when the boundary of a domain is advancing with a velocity proportional to the gradient of a field that satisfies the Laplace equation outside the domain. 
A characteristic feature of these processes is a strong instability of the interface motion: If the interface is an isoline of the harmonic field and the growth rate is proportional to the gradient of the field, small perturbations of the interface have a tendency to grow and eventually transform into fingers. At short wavelengths, the interface growth is stabilized by regularization mechanisms such as surface tension or kinetic undercooling, but the longer wavelengths are generally unstable. 
There are two main processes responsible for the pattern formation in these systems. The first is the screening between the nearby branches mediated by the harmonic field. As a result, longer branches tend to grow at an increased rate, whereas the growth of the shorter ones is impeded. The second process is tip splitting, when the branch bifurcates giving rise to a pair of daughter branches. The interplay of these two processes results in a highly ramified fractal structure of the advancing front.

\begin{figure*}[t]
\centering     
\includegraphics[width=0.9\textwidth]{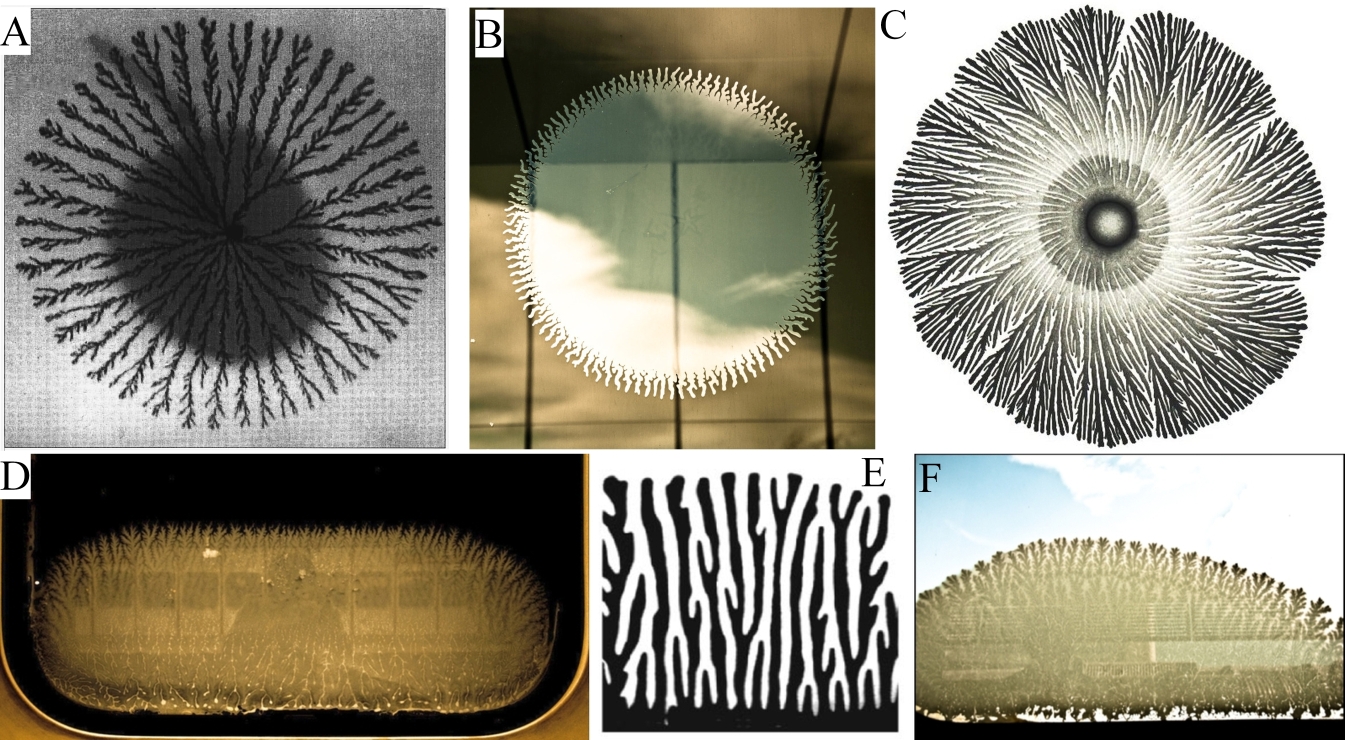}
\caption{Examples of unstable growth patterns with stable envelopes: (a) electrochemical deposition of zinc ions \cite{Lin1996}, (c) a colony of the tip-splitting morphotype bacteria of Paenibacillus dendritiformis \cite{benjacob1998}, (e) pattern generated in combustion in a Hele-Shaw cell \cite{Zik1998}, and (b), (d), and (f) the viscous fingering patterns in the windows of Vienna underground stations, created as the air invades the polyvinyl film separating the window panes.}
\label{comb}
\end{figure*}

In spite of such a strong instability of the interface, in many cases the emerging patterns show surprising regularities on a coarser scale, with a smooth envelope advancing in a stable way and forming a perfect circle (in radial geometry) or remaining planar (in rectangular geometry).  Examples of such patterns are presented in Fig.~\ref{comb}.  Arguably, the best known among them is the so-called dense-branched morphology observed in some of the electrodeposition experiments at increased voltages and electrolyte concentrations ~\cite{Grier1987,Garik1989,Barkey1991,Barkey1992,Grier1993,Fleury1994,Lin1996,Leger2000} [Fig.~\ref{comb}(a)]. However, similar regular envelopes have also been observed in bacterial colony growth~\cite{benjacob1998} [Fig.~\ref{comb}(c)], smoldering in quasi-two-dimensional (2D) systems~\cite{Zik1998,Zik1999} [Fig.~\ref{comb}(e)], or viscous fingering in a Hele-Shaw cell~\cite{benjacob1986,may1989,yeung1990,bischofberger2014,bischofberger2015}. Impressive examples of viscous fingering patterns with regular envelopes develop in layered window panes with imperfect sealing \cite{garcia1993,Ball2009},  which can be observed, for example, in some of the stations of Vienna's underground [Figs.~\ref{comb} 1(b), 1(d), 1(f)].

A number of different stabilizing mechanisms have been proposed over the years to serve as a theoretical explanation of this phenomenon. Grier \textit{et al.} \cite{Grier1987,Grier1993} argued that, in the context of electrodeposition, the growth is stabilized by the electrical potential drop across the filamentary pattern. A key element here is the anisotropic conductivity of the deposit, with the current flowing preferentially along the fingers. However, the anisotropic conductivity in the deposit can only stabilize the growth of the envelopes in circular geometry, but not in the planar geometry. To explain a stable growth in the latter setting, Lin and Grier \cite{Lin1996} invoked the effect of finite diffusion length. Namely, if the interface advances with velocity $v$, then beyond the lengthscale $l_d=D/v$ the fingers do not screen each other and  long-wavelength modes of the interface motion become stabilized. 
Other mechanisms  proposed to explain the stabilization of the envelope include the effects of electroconvection \cite{Fleury1994}, or the impact of a large concentration gradient near the interface, which can introduce an effective interfacial energy and the associated capillarity effects \cite{Barkey1991}.
In the context of viscous fingering, Couder \cite{Couder1988} suggested that the regular growth observed by Ben-Jacob \textit{et al.} \cite{benjacob1986} might be connected to the flexion of the plexiglass plates forming a Hele-Shaw cell in their experiments. The flexion makes the fingers move in  a  gap  of  varying  thickness,  stabilizing  the  extremity of  all  the  branches at a well-determined  position  of the  widening  gap. Overall, it seems likely that the stable growth forms such as those depicted in Fig.~\ref{comb} can be an example of equifinality, i.e., different combinations of processes or causes producing a similar form.

In this paper we propose a very general yet simple mechanism leading to the stabilization of the envelopes of the growing patterns. The only prerequisite for it to be applicable is that high growth velocity should trigger tip splitting of the fingers, which is a property shared by many pattern-forming systems \cite{Kassner,degregoria1985,meiburg1988,tabeling1988,kopf1988,Zhao1992,moore2002}. 
Thus, somewhat paradoxically, the regularity at a large scale is not despite but because  of a highly unstable behavior at a small scale.

\section{Thin-finger model}

For a theoretical description of a growing interface, we adopt a thin-finger model, in which the fingers are approximated by thin lines growing in response to the Laplacian field $\Psi({\bf r})$ ~\cite{Selander:1999,Carleson2002,Gubiec2008,Duran2010}. There are several advantages of such a model. First, it is analytically tractable and yields closed-form solutions in single- and two-finger cases. Second, 
it avoids the ultraviolet catastrophe at small wavelengths without the need to introduce a short-scale regularization such as surface tension. At the same time, the model preserves all the key features of the Laplacian growth such as long-range interaction between the fingers, which leads to their mutual screening. Models of this kind have been successfully used to simulate a number of pattern-forming processes with an underlying Laplacian field, such as the growth of the seepage channel networks~\cite{Devauchelle2012,Petroff2013,Cohen2015}, modeling of smoldering combustion~\cite{Gubiec2008}, growth of anisotropic viscous fingers~\cite{Pecelerowicz2014}, and diffusion-limited growth~\cite{hastings2001,gruzberg2004,Duran2010}. 

\begin{figure}[hb]
\centering     
\includegraphics[width=0.44\textwidth]{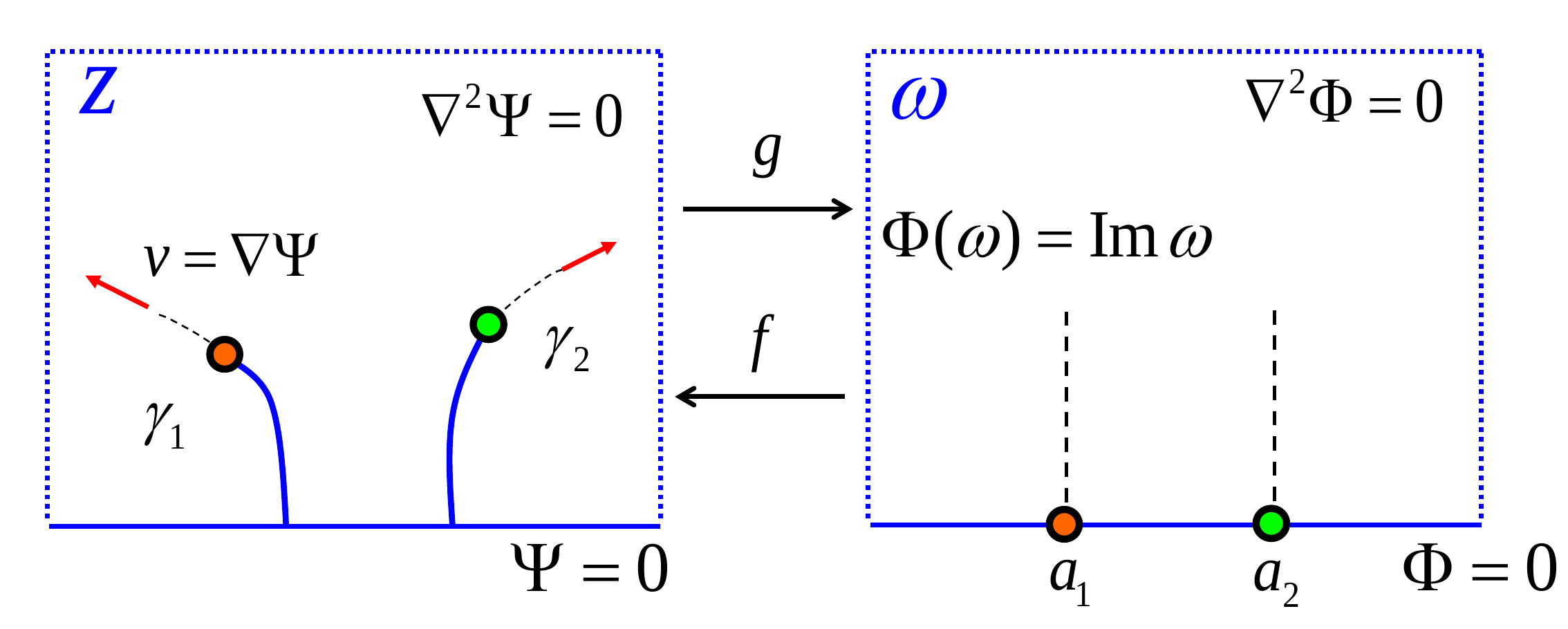}
\caption{Mapping $g_t$ of the exterior of the
fingers onto the empty system ($\omega$ plane). The images of the tips
$\gamma_i(t)$ are located on the real line at the points $a_i(t)$. The gradient lines of the Laplacian field
in the $z$ plane are mapped onto the vertical lines in the $\omega$ plane. At a given moment of time, the fingers grow along the gradient lines, the images of which pass through the points $a_i$.
}
\label{fig2}
\end{figure}

However, there is a consequence of the simplification: Since the finger is assumed to be infinitely thin, there is a singularity in a field gradient at its tip. Namely, at a small distance $r$ from the tip of the $i$th finger, the field takes the form
\begin{eqnarray}
\Psi_i({\bf r},t) =C_i(t) \sqrt{r} \cos (\theta /2),
\label{pre}
\end{eqnarray}
where the coefficients $C_i(t)$ depend on the lengths and shapes of all the fingers. In the above, the origin of coordinates is located at the tip of the finger and the polar axis is directed along it. The pressure gradient will then have $r^{-1/2}$ singularity. To address this issue, following Derrida and Hakim \cite{DerridaHakim:1992}, we
introduce a small circle of radius $r_0$ around the tip and define the finger growth rate as the integral of the field gradient over the circle
\begin{equation}
v_i(t) = \oint \hat{n} \cdot \nabla \Psi({\bf r},t) \  ds \ = 2 \sqrt{r_0} \ C_i(t).
\label{eq:v_int}
\end{equation}
The parameter $r_0$ should be of the order of the finger width; its exact value does not influence the dynamics as long as we assume it
to be the same for each finger. In such a case, the factor $2 \sqrt{r_0}$ may be absorbed
into the definition of time, and we subsequently take  $v_i(t)$ equal to $C_i(t)$.

Because of the quasi-2D geometry of the system, the Laplace equation is conveniently solved by the conformal mapping techniques~\cite{bazant2005}. To this end, one finds a mapping $g_t(z=x+iy)$ that transforms the region outside the fingers onto the empty system ($\omega$ plane in Fig.~\ref{fig2}). The solution of the Laplace equation in the $\omega$ plane, vanishing on the real axis, is simply $\Phi(\omega)=\beta \ \text{Im}(\omega)$, with the coefficient $\beta=|\nabla \Psi_{\infty}|$ prescribing the value of the field gradient at infinity.   This yields the potential of the form $\Psi(z)=\beta \ \text{Im}[g(z)]$ when transformed back onto the original domain. The description of the system in terms of $g_t$ is remarkably convenient, as $g_t$ can be shown to obey a first-order ordinary differential equation (deterministic Loewner equation), which represents a considerable simplification in comparison to the partial differential equation describing the boundary evolution.

\begin{figure}
\centering     
\includegraphics[width=0.5\textwidth]{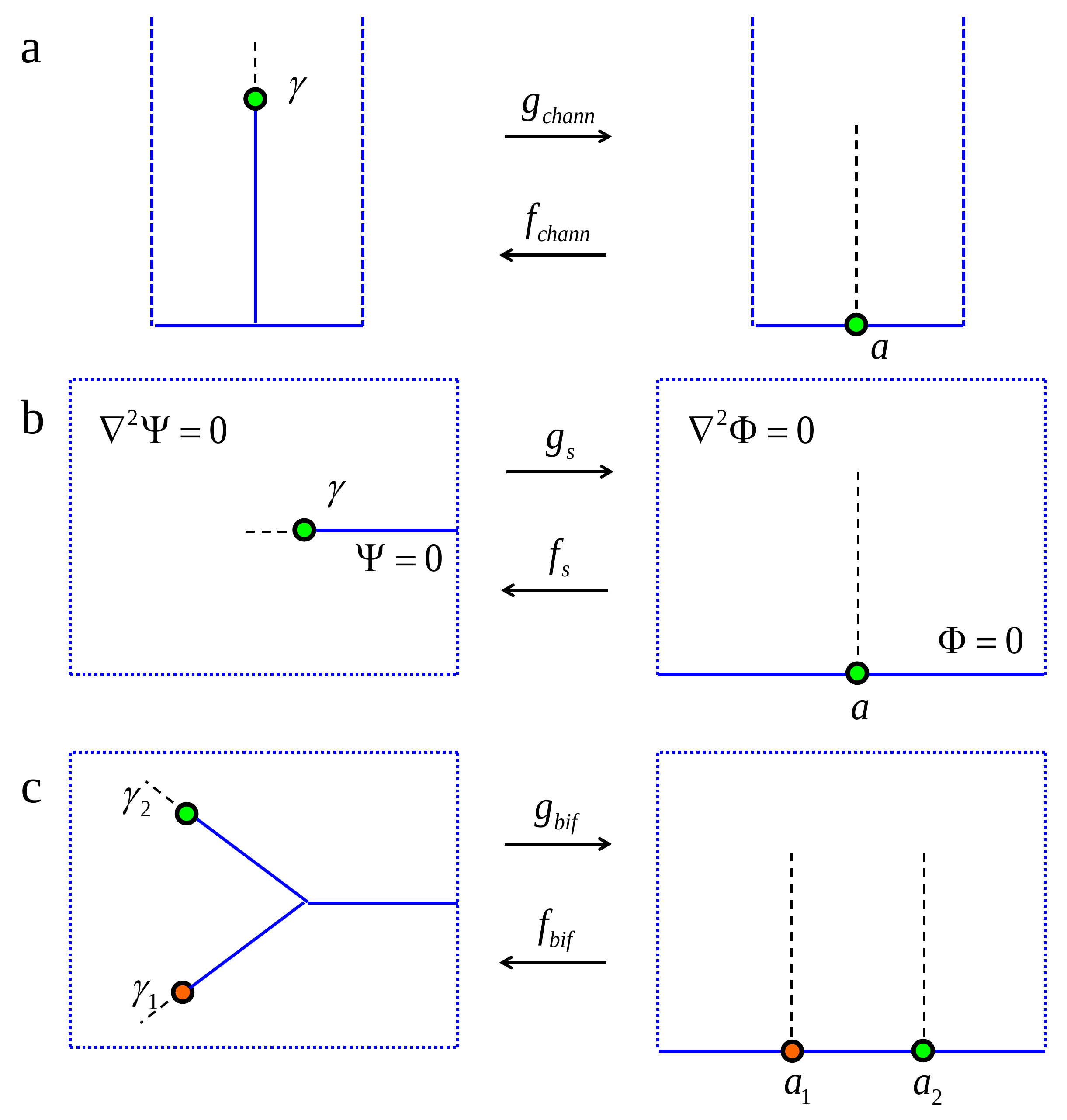}
\caption{Three conformal mappings used in the text: (a) mapping from the interior of an empty channel to the region outside of a single finger and mapping from an upper half space to the region around the tip of a finger just (b) before the bifurcation and (c) after the bifurcation.}
\label{mappy}
\end{figure}

The exact form of Loewner equation depends on the shape of the domain in which the growth takes place~\cite{Lawler:2005,Bauer:2006}. For example, for the growth of thin fingers in the channel with periodic boundary conditions, it reads \cite{Gubiec2008}
\begin{equation}
\dot{g}_{t} = \sum_{i=1}^{n} d_i(t) \frac{\pi}{W} \cot\left(\frac{\pi}{W}
[g_{t}(z)-a_i(t)] \right),
\label{loewner}
\end{equation}
with $W$ standing for the width of the channel and the initial condition $g_0(z) = z$ corresponding to the empty space with no fingers. Loewner equations for other geometries are given in the Appendix B. Note that the poles of the right-hand side of Eq.~\eqref{loewner} are located at the images of the tips $a_i(t)=g_t(\gamma_i)$ (cf. Fig.~\ref{fig2}).
The functions $d_i(t)$ are the so-called growth factors, controlling the speed with which the fingers are growing. By Taylor expanding the inverse mapping, $f_t=g_t^{-1}$ around $a_i(t)$ the exact relation between $d_i(t)$ and $v_i(t)$ can be shown to be \cite{Carleson2002,Gubiec2008} $d_i(t)={v_i(t)}/{|f^{\prime\prime}_t(a_i(t))|}$. On the other hand, the field amplitudes $C_i(t)$ in \eqref{pre} can also be expressed in terms of the conformal mapping $f_t$ \cite{Gubiec2008} as $C_i(t)=\sqrt{{2}/{|f^{\prime\prime}_t(a_i(t))|}}$. Hence, eventually,
\begin{equation}
 v_i(t) = \sqrt{2} |\nabla \Psi_{\infty}| \left| f''_{t}(a_i(t))\right|^{-1/2}.
 \label{eq:dlaplace}
\end{equation}
and
\begin{equation}
 d_i(t) = \sqrt{2} |\nabla \Psi_{\infty}| \left| f''_{t}(a_i(t))\right|^{-3/2}.
 \label{eq:dlaplace2}
\end{equation}
On the other hand, the evolving pole positions $a_i(t)$ in the Loewner equation \eqref{loewner} control the shape of the growing fingers. If the latter grow along the field lines, then the pole positions need to obey \cite{Cohen2015}
\begin{equation}
f^{\prime\prime\prime}_t(a_i(t))=0, \ \ i=1,N.
\label{polem}
\end{equation}
In a periodic channel, Eq.~\eqref{polem} is fulfilled provided that the poles move according to \cite{Gubiec2008}
\begin{equation}
\dot{a}_j = \sum_{i \neq j} d_i \frac{\pi}{W} \cot\left(\frac{\pi}{W}
(a_j-a_i) \right). \label{lcylpoles}
\end{equation}

\section{Tip splitting}
\label{ts}

Experimental and numerical observations on Laplacian growth systems suggest that, at least in some of the cases, tip splitting is triggered as the propagation velocity of a finger exceeds some critical velocity $v_{c}$ \cite{degregoria1985,meiburg1988,tabeling1988,kopf1988,Zhao1992,moore2002}.
On the theoretical side, it has been shown that viscous fingers are linearly stable up to the critical  propagation velocity at which they tip split~\cite{bensimon1986b,kessler1986b}. The exact value of this velocity depends on the amount of noise present in the system \cite{bensimon1986b,kessler1986b,tabeling1988}.

Within the thin-finger model, tip splitting corresponds to the creation of a pair of poles out of a single one. Thus we will assume that whenever $v_i \geq v_c$  finger $i$ will be split into $i_1$ and $i_2$ with
 \[\begin{array}{l}
{{a}_{i_{1}}}({t_0}) = {a_i}({t_0}) + \varepsilon \\
{{a}_{i_{2}}}({t_0}) = {a_i}({t_0}) - \varepsilon 
\end{array}\]
where $\varepsilon$  is an infinitesimal positive constant. The shape of the finger in the vicinity of the bifurcation can be obtained by noting that the conformal mapping $f_t$ that maps the upper half plane to the region outside of the single symmetric bifurcation with the opening angle $\alpha$ [Fig.~\ref{mappy}(c)] reads \cite{Carleson2002}
\begin{equation}\label{eq:bifurcation_map}
f_{bif}( \omega ) =  \omega^{\alpha/\pi} \left(\omega-\sqrt{\frac{2 \pi}{\alpha}} a_1 \right)^{1 - \alpha/2\pi} \left(\omega+\sqrt{\frac{2 \pi}{\alpha}} a_1 \right)^{1 - \alpha/2\pi},
\end{equation}
where $a_1=-a_2$ is the pole position. Imposing \eqref{polem} gives a universal bifurcation angle $\alpha=2/5 \pi$ as noted in a number of previous studies \cite{hastings2001,Carleson2002,Cohen2015}. 
 The ratio of velocities after and before the bifurcation is given by $\chi=\sqrt{\left| \frac{f_{s}^{\prime\prime}(0)}{f_{bif}^{\prime\prime}(a_1)} \right|}=2^{-3/10} \approx 0.812$, where $f_s(\omega)=\omega^2$ is the respective mapping for a mother finger before the bifurcation [cf. Fig.~\ref{mappy}(b)]. 
Hence, during the evolution, the velocities of the active fingers oscillate between $v_s=0.812 v_c$ and $v_c$.
\begin{figure}[h]
\begin{center}
\includegraphics[width=7cm]{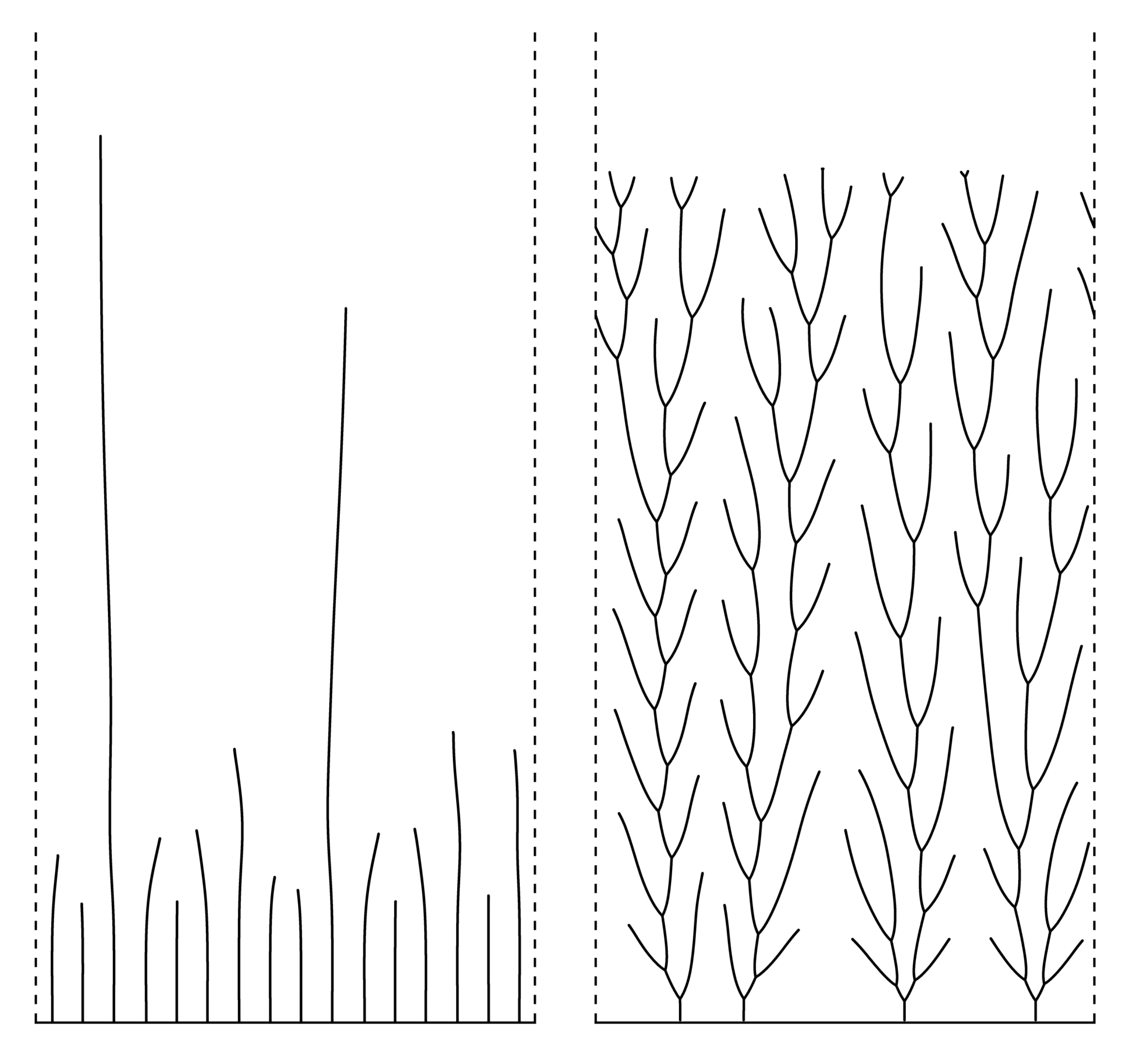}
\caption{Growth patterns in a channel geometry with (right) and without (left) tip splitting. In all the simulations reported in this paper $v_{c}=\sqrt{6}$.}
\label{inten}
\end{center}
\end{figure}

\begin{figure}[h]
\begin{center}
\includegraphics[width=0.35\textwidth]{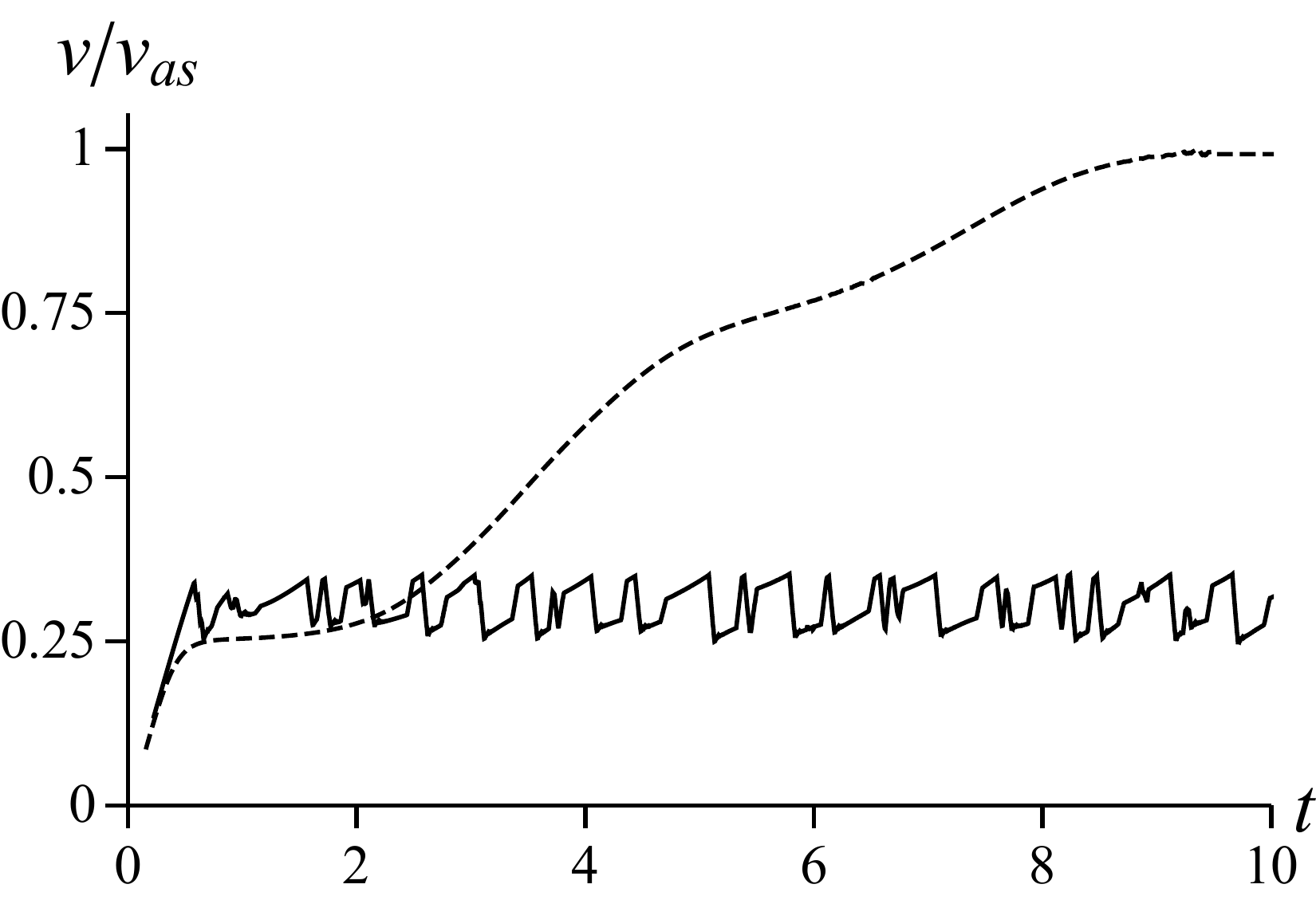}
\caption{Front advancement velocity (equal to the growth velocity of the longest finger) vs time for the system without tip splitting (dashed) and with tip spitting (solid).}
\label{frontspeed}
\end{center}
\end{figure}

\section{Results}

Figure ~\ref{inten} presents a typical growth pattern in a channel geometry obtained using the above model with and without tip splitting. There is a stark contrast between the two cases: Without tip splitting the main process controlling the evolution of the pattern is screening between the neighboring fingers.
The longer fingers collect an increasingly larger portion of the total flow and thus grow with an increasing velocity at the expense of the shorter ones. The distance between the active (growing) fingers constantly increases. Finally, when it becomes comparable to the system width, a single winning finger remains.  Its asymptotic growth velocity can be calculated by noting that a conformal transformation that maps the interior of an empty channel to the region outside a single finger is [cf. Fig.~\ref{mappy}(a)]
\begin{multline}
f_{chann} (z) = \!\! \frac{W}{\pi} \arcsin \left(     \sin^2 (\frac{ \pi}{W} z) \right.
\cosh^2 (\frac{\pi}{W} H(t))  \\  \left.  -\sinh^2 (\frac{ \pi}{W} H(t) )
\right)^{1/2},
\label{mapa1}
\end{multline}
where $H(t)$ is the height of the finger at a given moment of time. Using \eqref{eq:dlaplace} one gets the growth velocity of a finger as
\begin{equation}
v(H)= \sqrt{2\frac{W}{\pi}} \left(\coth \left( \frac{\pi}{2} H \right) \right)^{-1/2} |\nabla \Psi_{\infty}|
\label{oj1}
\end{equation}
which asymptotically converges to
\begin{equation}
v_{as}=\lim_{H \rightarrow \infty} v(H) =   \sqrt{\frac{2W}{\pi}} |\nabla \Psi_{\infty}| 
\label{vas}
\end{equation}
On the other hand, when tip splitting is allowed, the screening between the fingers is compensated by the creation of new ones and the system quickly reaches a stationary state, with a constant average number of fingers across the width. The density of the fingers in such a situation can be estimated based on Eq.~\eqref{vas} by noting that each of $N$ fingers is growing effectively in a strip of width $w_N=W/N$, hence 
\begin{equation}
v \approx \sqrt{\frac{2W}{N\pi}} |\nabla \Psi_{\infty}| = \sqrt{\frac{2}{n\pi}} |\nabla \Psi_{\infty}|
\label{b12}
\end{equation}
where $n=N/W$ is the density of the fingers in a given place along the envelope. 
On the other hand, the active fingers are always on the edge of splitting, hence $v \approx v_c$. This leads to the following estimate of the finger density
\begin{equation}
n \approx \frac{2}{\pi} \frac{|\nabla \Psi_{\infty}|^2}{v_c^2} 
\label{skal}
\end{equation}

\begin{figure}[h]
\begin{center}
\includegraphics[width=7cm]{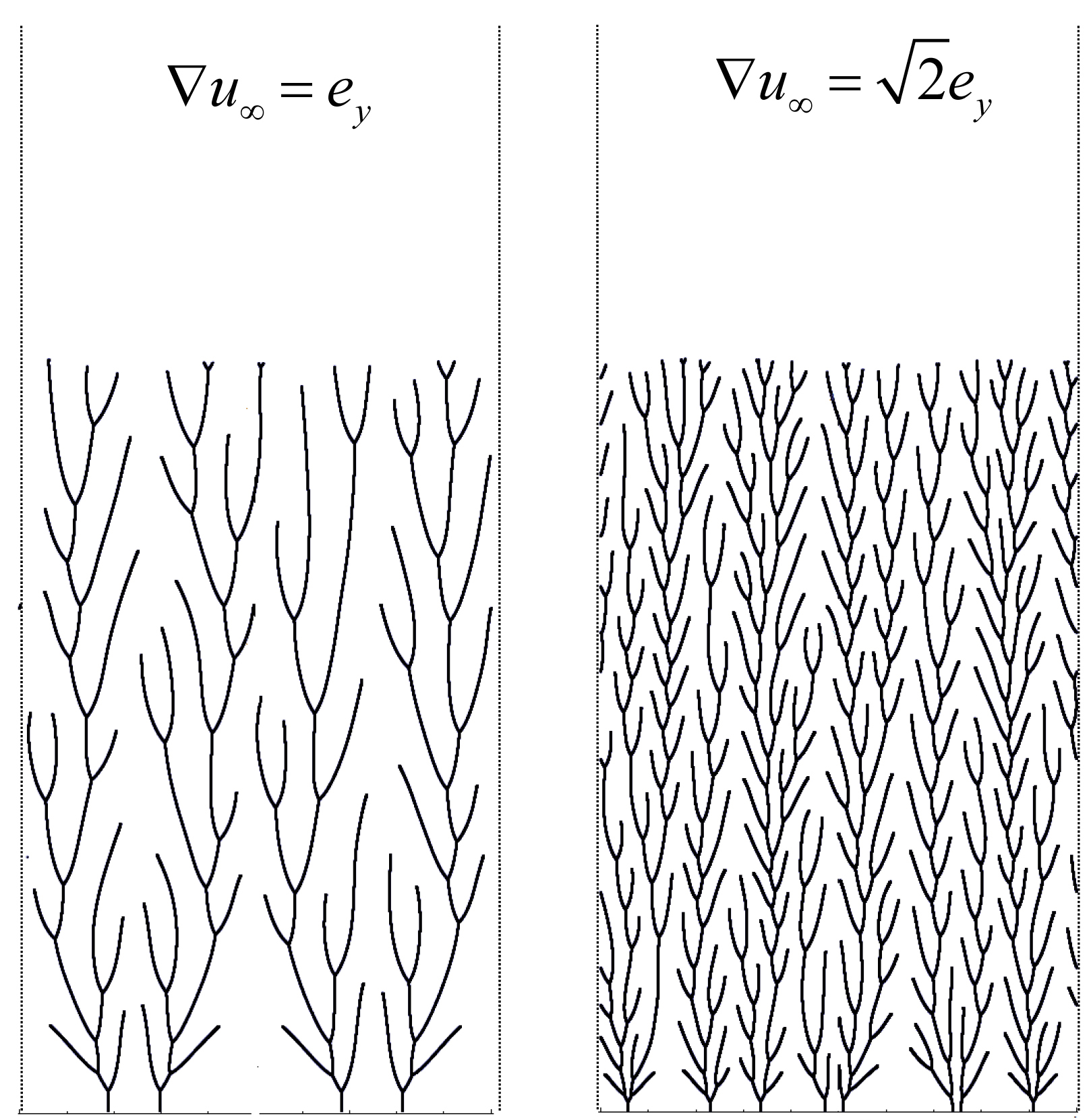}
\caption{Growth patterns in a rectangular geometry for two different intensities of a driving field captured at the same moment in time. The complete evolution of the pattern is shown in a corresponding movie in the Supplementary Material~\cite{Note1}.}
\label{dwa}
\end{center}
\end{figure}

The differences between the case with and without the tip splitting are further elucidated by the analysis of the time dependence of the front advancement speed, defined as the velocity of the longest finger. As observed in Fig.~\ref{frontspeed}, without tip splitting the front velocity, after an initial sharp rise, saturates near the value of $0.25 v_{as}$. This is the moment when all of the 16 fingers in the system  [cf. Fig.~\ref{inten}(a)] are of a similar height, each growing with a velocity of $1/\sqrt{16} v_{as}$, in accord with the analysis presented above. Then, however, the fingers begin to screen each other off and the number of active ones decreases, which is accompanied by a respective rise in the front speed (following the rule $v \sim 1/\sqrt{N_{act}}$). The second pronounced plateau in $v(t)$ dependence corresponds to the situation when only two fingers remain and $v \approx 1/\sqrt{2} v_{as}$. Finally, a single active finger is left in the system and its speed reaches $v_{as}$. 

A markedly different situation is encountered in the case with tip splitting. Here, after an initial sharp rise of the growth velocity, the system reaches a steady state, where the speed of the leading finger oscillates between $v_c$ and $v_s \approx 0.8 v_c$, as elucidated in Sec.~\ref{ts}.

Equation~\eqref{skal} suggests that the density of the fingers scales quadratically with the driving current. This is further confirmed by the analysis of Fig.~\ref{dwa}, which indeed shows that the average number of fingers per width of the system increases approximately twofold, as the far-field gradient is increased by a factor of $\sqrt{2}$. At the same time the advancement velocity of the pattern remains constant (both panels of Fig.~\ref{dwa} present the patterns captured at the same moment in time). 

The above considerations elucidate the mechanism of the stable movement of the envelope, as observed in Figs.~\ref{inten}-\ref{dwa}. Namely, if the tip splitting is absent, an increased flux $J$ impinging at a finger leads to its faster growth ($v \sim J$) and screening of its neighbors. On the other hand, in the presence of tip splitting, the increased flux in a given point at the boundary results in an increased frequency of the splitting events. As a result, the relaxation of the flux proceeds through an increased density of the fingers ($n \sim J^2$), but the envelope of the pattern  moves steadily ($v \sim v_c$).

\section{Inhomogeneous systems}

The periodic channel considered in the previous section is characterized by a high degree of homogeneity: The emerging pattern is uniform, except for the fluctuations connected with the splitting-screening cycle. To go beyond this case, in this section we consider the growth in two different inhomogeneous systems. 

\begin{figure}[h]
\begin{center}
\includegraphics[width=7cm]{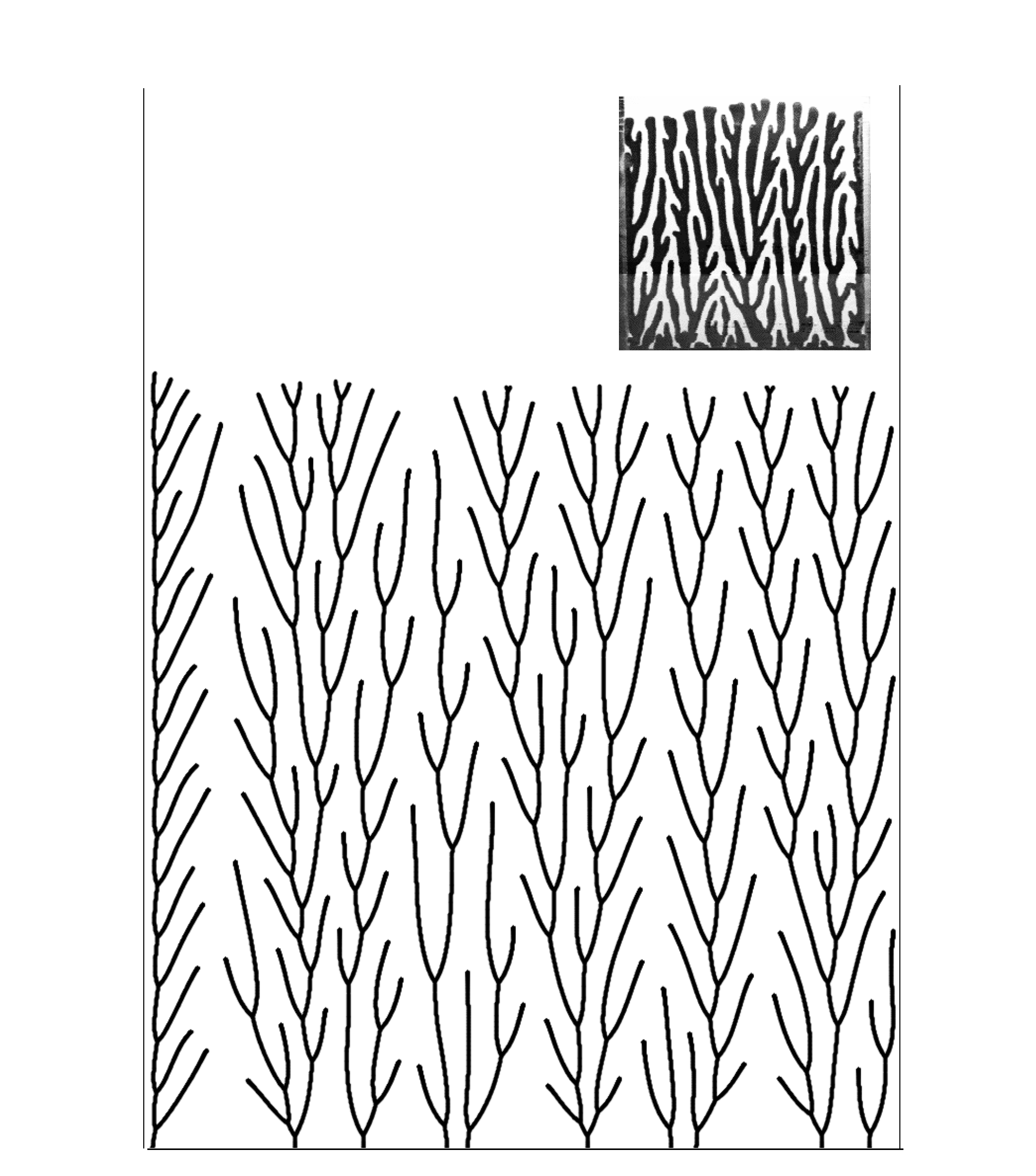}
\caption{Growth patterns in a channel with reflecting walls with a characteristic asymmetric structure at the left wall. Similar structures are observed near the walls in the combustion experiments of Zik and Moses \cite{Zik1999} (inset).}
\label{channref}
\end{center}
\end{figure}

First, let us analyze the growth in the channel with reflective side walls. As observed in Fig.~\ref{channref}, far from the wall the pattern is similar to that in the periodic channel. Near the walls, however, the fingers look qualitatively different. 

If the finger grows in close proximity to a reflecting wall, it strongly interacts with its image behind the wall.  In the absence of other fingers in the system, this interaction would repel the finger from the wall at an angle of $\pi/10$ with respect to the vertical. This is because the angle between two interacting lines in a half plane tends to $\pi/5$ in the long-time limit \cite{Gubiec2008}. If such a slanted finger splits, the two daughter branches are now not moving symmetrically with respect to the vertical axis and the one closer to the wall wins. The process then repeats itself,  finally resulting in an almost vertical finger growing close to the wall occasionally releasing side branches  towards the bulk of the system. Structures of this kind are commonly observed in experimental systems (cf. the inset of Fig.~\ref{channref}). Importantly, these structures arise only if the main branch of the finger grows sufficiently close to the wall so that the interaction with the image is stronger than that with the sibling branch. For larger distances between the main branch and the wall, the effect ceases to be present, as it is the case near the right wall of the system in Fig.~\ref{channref}.

Finally, let us look at the growth taking place in a half-plane bounded by an isopotential line with a constant field gradient at infinity. Starting from a single seed, we observe the growth of a tree-like structure, with intense bifurcations. Importantly, in this case the field gradient is nonuniform along the boundary of the structure, with the highest value at its top. Nevertheless, the 
envelope forms a perfect semicircle, expanding uniformly and preserving its shape in time. The density of the fingers (and frequency of splittings) is larger at the top as more flux needs to be absorbed there. This effect can also be observed in the window patterns of Figs.~\ref{comb}(d) and ~\ref{comb}(f) with the average distance between the fingers on the side of the structures significantly larger than along the top. 
 
More quantitatively, the Laplace potential around a grounded semicircle of radius $r$ is
\begin{equation}
\Psi(z)=\text{Im}(z + \frac{r^{2}}{z})
\end{equation}
where a unit gradient at infinity has been assumed. The field gradient at the surface is then
\begin{equation}
|\nabla \Psi|=2 \sin (\theta),
\end{equation}
where $\theta$ is an angle from a real axis. Based on considerations similar to before, the density of the active fingers is expected to behave as
\begin{equation}
n \approx \frac{2}{\pi} \frac{4\sin(\theta)^{2}}{v_c^2},
\end{equation}
with the total number of fingers scaling linearly with $r$. Figure~\ref{distro} presents the cumulative distribution of the fingers $N(r,\theta)=\int_0^{\theta} n(\theta^{\prime}) r d \theta^{\prime}$ measured in the simulations compared with the theoretical prediction:
\begin{equation}
N(r,\theta) = \int\limits_0^\theta  n(\theta^{\prime}) r\text{d}\theta^{\prime}  = \frac{{4r}}{{\pi v_c^2}}\left( \theta  - \sin \theta \cos \theta  \right).
\label{pred}
\end{equation}
Good agreement between the two shows that the simple model presented here indeed captures the key elements of the dynamics of these systems. Note that the comparison with Eq.~\eqref{skal} requires counting not all of the fingers present in the pattern, but only the active ones, i.e., the ones that are growing and would eventually split. This can be assessed based on the velocity of the fingers: The active ones would invariably have velocities between $v_s$ and $v_c$ (see Sec.~\ref{ts}), whereas the dying branches move with much lower speeds.  However, the data in Fig.~\ref{distro} show that the distribution of the total number of fingers is of a very similar form to $N(\theta)$ above, only rescaled by a factor close to $2$. This can be rationalized by noting that the strongest screening interactions arise between the neighboring fingers, very often the daughter branches emerging from the same mother finger. As a result of such mutual screening, every second finger, on average, loses the competition and dies. 

Returning to the evolution of the shape of the system, it is worth noting that if the envelope were evolving according to the standard Laplacian growth law $v \sim \nabla \Psi$, the semicircle would not preserve its shape but instead would transform into a half-oval of eccentricity increasing in time, since the field gradient is highest at the top of the structure (for $\theta = \pi/2$) and then tapers towards the sides. Thus, in the case of tip splitting systems the naive upscaling of the growth law from the microscopic (single-finger) scale towards the macroscopic (envelope) scale does not lead to the correct growth law. In fact, the motion of the envelope is rather governed by the relation
\begin{equation}
v_n = \text{min} \left( {v_c,\sqrt{\frac{2}{n\pi}}{{\left( {\nabla \Psi} \right)}_\perp}} \right),
\end{equation}
where the subscript $\perp$ stands for the velocity component normal to the envelope and we have used Eq.~\eqref{b12} linking the propagation velocity and the local density of the fingers.

\begin{figure}
\centering     
\includegraphics[width=0.5\textwidth]{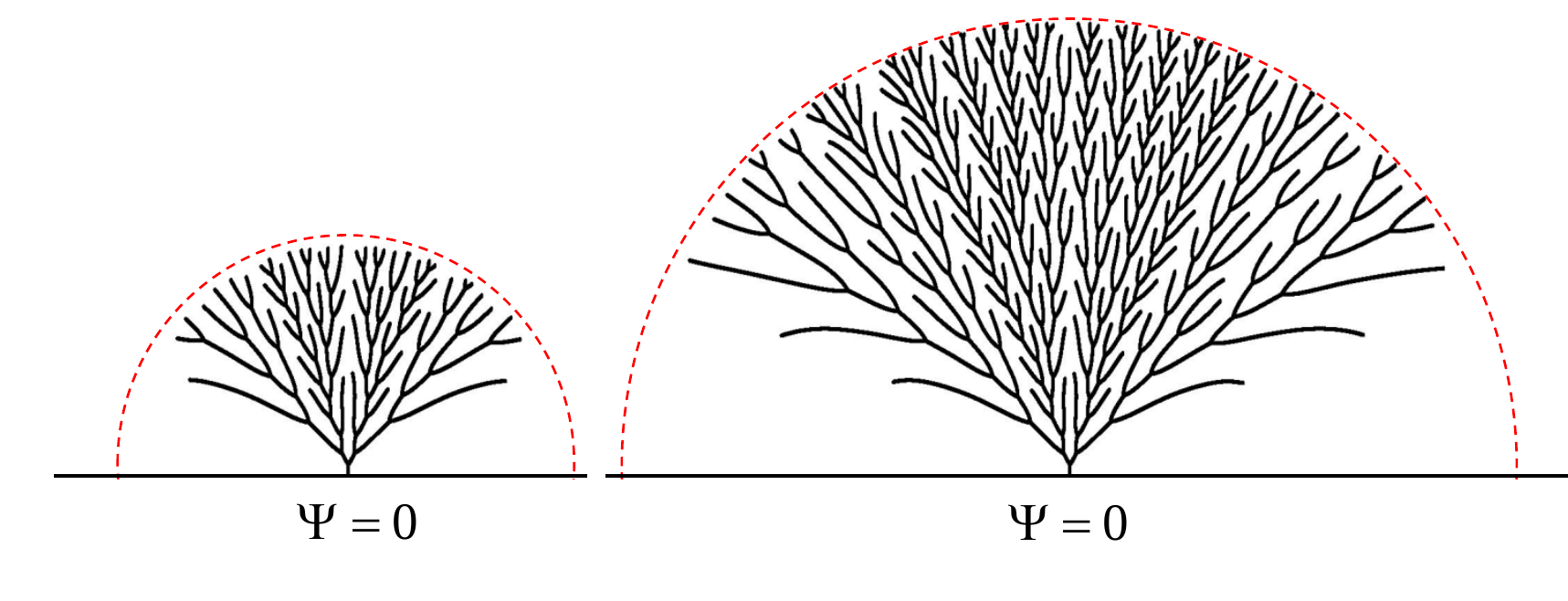}
\caption{Evolution of the growing pattern in a half-plane geometry. A corresponding movie can be found in the Supplementary Material ~\cite{Note2}.}
\label{tree}
\end{figure}

\begin{figure}[h]
\begin{center}
\includegraphics[width=0.4\textwidth]{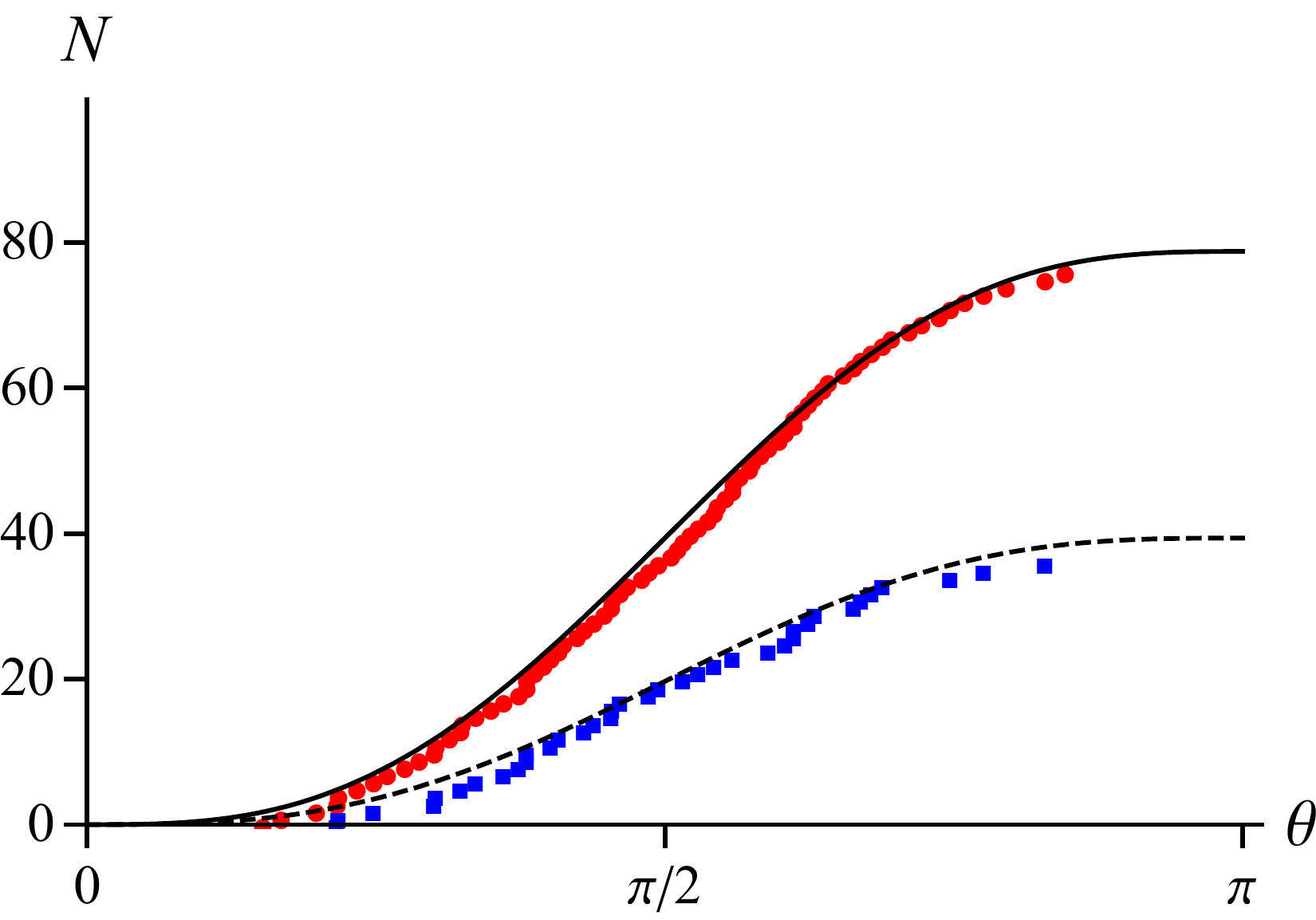}
\caption{Cumulative distribution of the fingers along the boundary of the Laplacian tree of Fig.~\ref{tree} counting all of the fingers (red circles) or only the active ones (blue squares). The dashed line marks the theoretical prediction of the number of active fingers [Eq.~\eqref{pred}] for $r=58$, whereas the solid line corresponds to twice the theoretical prediction.}
\label{distro}
\end{center}
\end{figure}

\section{Interaction of the envelopes}

\begin{figure}[ht]
\begin{center}
\includegraphics[width=0.5\textwidth]{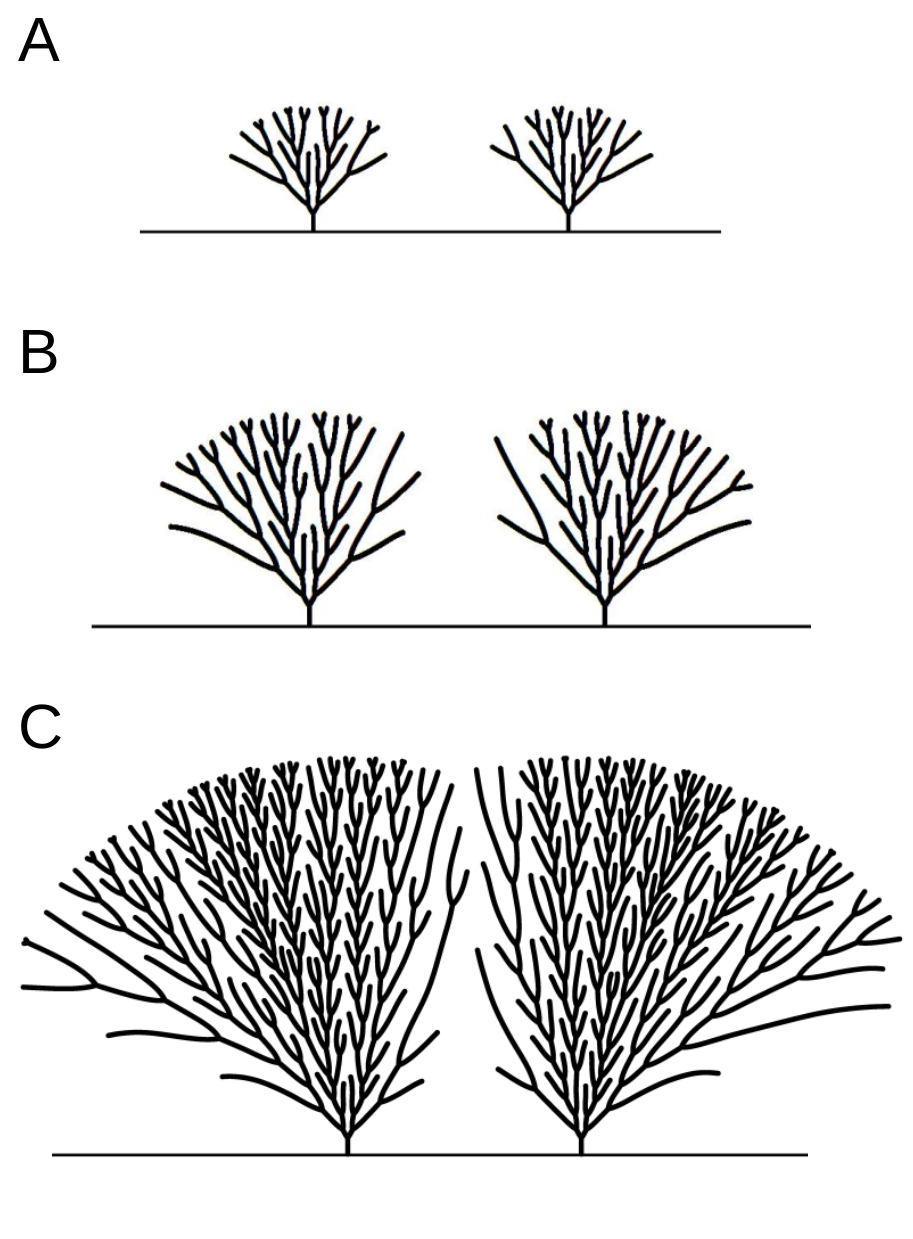}
\caption{Two Laplacian trees growing near each other in the half plane}
\label{porownanie}
\end{center}
\end{figure}

Finally, let us consider the interaction of two growing structures as their envelopes approach each other. 
In order to analyze it, we place two trees in the half-plane relatively far from each other so that initially they both grow freely and do not interact [Fig.~\ref{porownanie}(a)]. However, as they get closer to each other, the frequency of splitting on the interior sides drastically decreases. Nevertheless, the growth velocity remains constant [Fig.~\ref{porownanie}(b)]. As the region between the envelopes becomes strongly screened from both sides, the growth velocity drops below $v_c$ and the splitting stops. This results in a creation of a group of long non-splitting branches, which progressively slow down and finally stop growing. At the same time, the outer parts of both trees grow outward with a steady velocity. Finally, both trees merge together and the envelope of a resulting structure becomes semicircular itself [Fig.~\ref{porownanie}(c)]. 

\begin{figure*}[ht]
\begin{center}
\includegraphics[width=\textwidth]{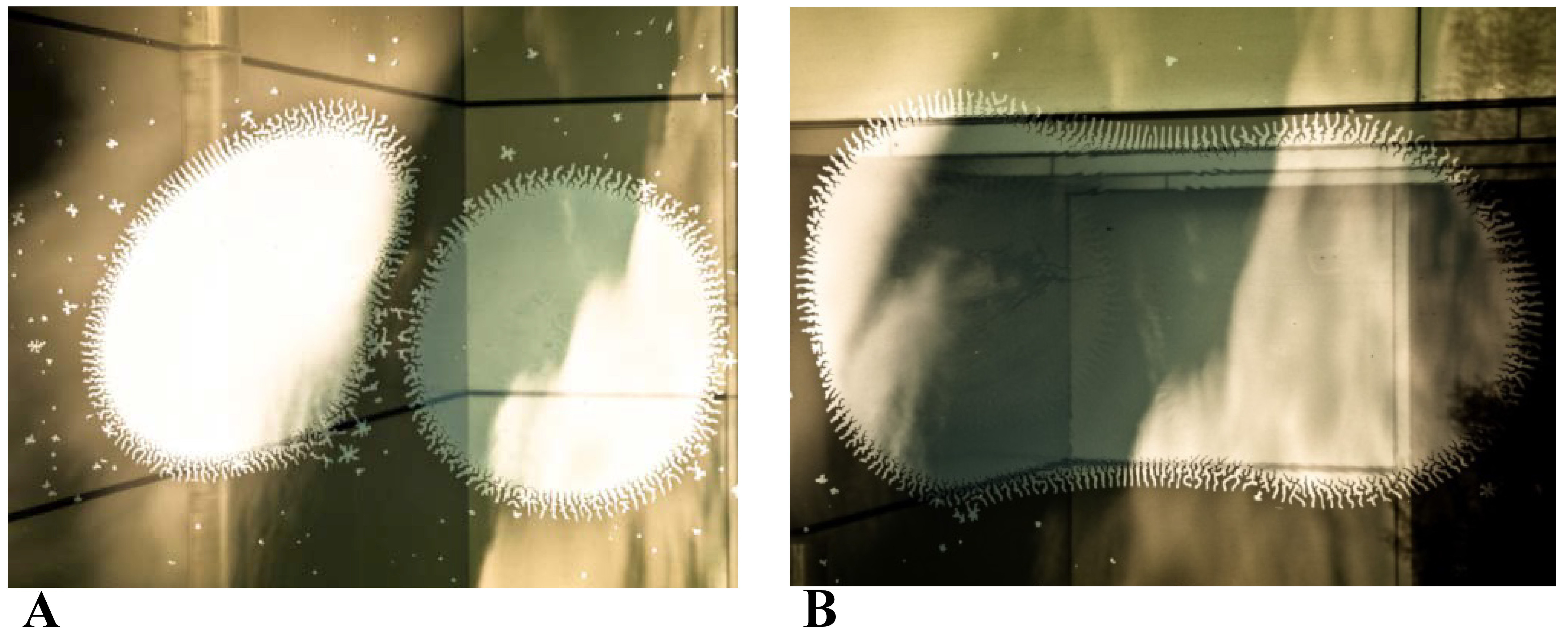}
\caption{Interacting viscous fingering patterns in  Vienna U-Bahn windows.}
\label{porownanie2}
\end{center}
\end{figure*}

Analogous dynamics can be observed in the viscous fingering patterns in Vienna U-Bahn windows.  Individual patterns formed in the central parts of the window, far from the neighbors, are highly circular [cf. Fig.~\ref{comb}(b)]. However, if two such patterns form close to each other [Fig.~\ref{porownanie2}(a)], their neighboring sides flatten out, until they merge [Fig.~\ref{porownanie2}(b)].

\section{Summary}

In this paper, we have presented a simple mechanism that leads to the stabilization of the envelope motion in the fingered growth system. In the model, a local increase in the field gradient at the boundary of a growing structure does not result in an accelerated growth. Instead, the frequency of splitting events is increased, which leads to a higher density of the fingers. This ultimately stabilizes the global growth of the pattern as a whole by absorbing the excess flux without increasing the advancement velocity. There are two main prerequisites for the model to be applicable. First, the tip splitting needs to be associated with the threshold velocity of the advancement of the fingers. Second, the fingers should have a well-defined width, which should not change during the growth. An increase of the width might constitute an alternative way of relaxing the excessive flux, however, in the present model $r_0$ is assumed to be constant [cf. Eq.~(\ref{eq:v_int})].

\begin{acknowledgments}
{\small The authors benefited from discussions with Olivier Devauchelle and Joachim Mathiesen. We thank Agnieszka Budek for photographs of the patterns in Vienna U-Bahn windows [Fig. 1(b), 11(d), 1(f)] and Michal Ben Jacob for sharing the photo of bacterial colonies from the archives of late Professor Eshel Ben Jacob [Fig. 1(c)]. The image in Fig. 1(a) is courtesy of Professor David Grier, New York University, and the image in Fig 1(e) is courtesy of Professor Elisha Moses, Weizmann Institute, Israel. This work was supported by the National Science Centre (Poland) under Grant 2012/07/E/ST3/01734. We also thank the PL-Grid Infrastructure for computer resources. }
\end{acknowledgments}

    \appendix

\section{Numerical method}

Similarly to the previous works on the subject \cite{Gubiec2008,Kennedy2009}, we construct the solution of the Loewner equation by the composition of elementary slit mappings, each extending a given finger over the time interval $\tau$. For the growth in the cylinder, such a slit mapping reads
\begin{multline}\label{mabv}
\phi_i(z;\tau)=  \frac{W}{\pi}\arcsin \left( \tanh^2\left(\frac{\pi}{W} \sqrt{2\tau d_i} \right) \right. 
\\ \left. +\sin^2 \left(\frac{\pi}{W}(z-a_i) \right) \cosh^{-2}\left(\frac{\pi}{W} \sqrt{2\tau d_i}\right) \right)^{1/2}+a_i,
\end{multline}  
which is essentially an inverse of the mapping~\eqref{mapa1}. Since there are $n$ fingers, each timestep involves the composition of $n$ slit mappings $\phi_i$, each characterized by a corresponding position of the pole, $a_i$, and the growth factor $d_i$.
To calculate the growth factors the mapping $f_t$, inverse to $g_t$, is needed [cf. Eq.(\ref{eq:dlaplace2})].  This mapping can also be obtained by the composition of elementary mappings $\tilde{\phi}$, which are the inverses of slit mappings $\phi$, i.e., $\tilde{\phi}(\phi(z))=z$. 

Two points need to be mentioned here. First, the order of compositions of slit mappings corresponding to different fingers matters, since $\phi_j(\phi_i(z;\tau),\tau) - \phi_i(\phi_j(z;\tau),\tau) = O(\tau^2)$. To prevent the appearance of cumulative systematic error, we randomize the order in which slit mappings are applied in each time step. Second, special care needs to be taken while tracking the fingers just after the tip-splitting event, due to the presence of singularities in the pole evolution equation \eqref{lcylpoles} whenever $a_i \approx a_j$.  The direct composition of single-finger slit mappings leads then to significant errors. Instead, we apply then the V-shaped mapping \eqref{eq:bifurcation_map} [Fig.~\ref{mappy}(c)] with the opening angle $\alpha=2\pi/5$ between the branches.

    \section{Loewner equation for different geometries}

Below we summarize the form of the Loewner equation for different geometries considered in the present study. 

\subsection*{\bf Half plane}

In this case the domain in which the growth takes place is the upper half of the 
complex plane $\mathbb{H}=\{\omega \in \mathbb{C} | \text{Im}(\omega)>0\}$. The Laplace equation is solved 
in the region outside the fingers, $\Omega_{t}=\mathbb{H} \backslash K_{t}$. Here $K_t$ is the 
configuration of the branches in the physical plane at time $t$. 
The mapping $g_t$ takes $\Omega_{t}$ onto $\mathbb{H}$
\begin{equation} 
g_{t}:\Omega_{t}\rightarrow \mathbb{H}
\end{equation}
with the normalization 
\begin{equation}
g_{t}(z) \to z + O(1/z) \ \ \text{as \ \ \ }  z \rightarrow\infty
\label{norm}
\end{equation}
The Loewner equation in this case reads \cite{Popova1954}
\begin{equation}
\dot{g}_{t}(z) = \sum_{i=1}^{n}\frac{d_{i}(t)}{g_{t}(z) - a_{i}(t)}
\end{equation}
If the fingers are to grow along the field lines, the pole positions need to obey \cite{Carleson2002}
\begin{equation}
\dot{a}_{j}(t) = \sum_{i=1,i\neq j}^{n}\frac{d_{i}(t)}{a_{j}(t) - a_{i}(t)}
\end{equation}
Finally, an elementary slit mapping for this geometry reads
\begin{equation}
	\phi_i(z;\tau)=\sqrt{(z-a_i)^{2}+2\tau d_i}+a_i.
\end{equation}

\subsection*{\bf Cylinder (channel with periodic boundary conditions)}

The domain in which the growth takes place is 
\begin{equation}
	\mathds{P}=\left\{ z=x+iy\in \mathds{C}: y > 0 , x \in \ [-W/2,W/2[ \
\right\},
\label{ka1}
\end{equation}
with the Dirichlet boundary condition for the harmonic potential on both the fingers and the bottom wall $[-1,1[$ and periodic boundary conditions at the lateral sides $\Psi(x+W,y)=\Psi(x,y)$, which makes the system topologically equivalent to the surface of a {semi-infinite} cylinder. The mapping $g_t$ takes $\Omega_{t}=\mathds{P} \backslash K_{t}$ onto $\mathds{P}$
\begin{equation} 
g_{t}:\Omega_{t}\rightarrow \mathds{P}
\end{equation}
with the normalization 
\begin{equation}
g_{t}(z) \to z + O(1) \ \ \text{as \ \ \ }  z \rightarrow \infty
\label{norm2}
\end{equation}
The Loewner equation in this case reads \cite{Gubiec2008}
\begin{equation}
\dot{g}_{t} = \frac{\pi}{W} \sum_{i=1}^{n} d_i \frac{\pi}{W} \cot\left(\frac{\pi}{W}
(g_{t}-a_i) \right),
\label{ncyl}
\end{equation}
whereas the equation of motion of the poles is
\begin{equation}
\dot{a}_j = \frac{\pi}{W} \sum^{n}_{ \stackrel{i=1}{i \neq j} } d_i  \cot\left(\frac{\pi}{W}
(a_j-a_i) \right). 
\end{equation}
An elementary slit mapping for this geometry is given by Eq.~\eqref{mabv}.

\subsection*{\bf Channel with reflecting boundary conditions}

The domain in which the growth takes place is again  $\mathds{P}$ defined in \eqref{ka1}, but this time with Neumann boundary conditions $\frac{\partial \Psi}{\partial x} = 0$ at the lateral sides.  
The Loewner equation in this case reads \cite{Gubiec2008}
\begin{equation}
		 \dot{g}_{t} =\frac{\pi}{W}  \sum_{i=1}^{n} d_{i}
\frac{\cos\left(\frac{\pi}{W} g_{t} \right)}{\sin\left(\frac{\pi}{W} g_{t} \right) - \sin
\left(\frac{\pi}{W}a_{i}\right)}.
\label{Loewnern}
\end{equation}
whereas the condition for the motion of the poles is
\begin{equation}
		 \dot{a}_{j} = - \frac{\pi}{2W} d_{j}
\tan\left(\frac{\pi}{W}a_{j}\right)
 + \frac{\pi}{W}\sum^{n}_{ \stackrel{i=1}{i \neq j} } d_{i}  \frac{\cos\left(\frac{\pi}{
W} a_{j} \right)}{\sin
\left(\frac{\pi}{W} a_{j} \right) - \sin\left(\frac{\pi}{W}a_{i}\right)}.
\label{eq:aprimP_n}
\end{equation} 
An elementary slit mapping for this geometry has been derived in Ref.~\cite{Gubiec2008} [Eqs.~(28)-(30)]. The final expression is somewhat lengthy and thus we do not reproduce it here.

\bibliographystyle{apsrev4-1}


%

\end{document}